# The Unreasonable Effectiveness of Deep Learning in Artificial Intelligence

Terrence J. Sejnowski[a,b,*]

[a] Computational Neurobiology Laboratory, Salk Institute, La Jolla, CA, 92037 USA

[b] Division of Biological Sciences, University of California San Diego, La Jolla, California 92093 USA

**Abstract:**

Deep learning networks have been trained to recognize speech, caption photographs and translate text between languages at high levels of performance. Although applications of deep learning networks to real world problems have become ubiquitous, our understanding of why they are so effective is lacking. These empirical results should not be possible according to sample complexity in statistics and non-convex optimization theory. However, paradoxes in the training and effectiveness of deep learning networks are being investigated and insights are being found in the geometry of high-dimensional spaces. A mathematical theory of deep learning would illuminate how they function, allow us to assess the strengths and weaknesses of different network architectures and lead to major improvements. Deep learning has provided natural ways for humans to communicate with digital devices and is foundational for building artificial general intelligence. Deep learning was inspired by the architecture of the cerebral cortex and insights into autonomy and general intelligence may be found in other brain regions that are essential for planning and survival, but major breakthroughs will be needed to achieve these goals.

* Correspondence: terry@salk.edu

In 1884, Edwin Abbott wrote "Flatland: A Romance of Many Dimensions" (1) (Fig. 1). This book was written as a satire on Victorian society, but it has endured because of its exploration of how dimensionality can change our intuitions about space. Flatland was a two-dimensional world inhabited by geometrical creatures. The mathematics of two dimensions was fully understood by these creatures, with circles being more perfect than triangles. In it a gentleman square has a dream about a sphere and wakes up to the possibility that his universe might be much larger than he or anyone in flatland could imagine. He was not able to convince anyone that this was possible and in the end he was imprisoned.

We can easily imagine adding another spatial dimension when going from a one- to a two-dimensional world and from a two- to a three-dimensional world. Lines can intersect themselves in two dimensions and sheets can fold back onto themselves in three dimensions. But imagining how a three-dimensional object can fold back on itself in a four-dimensional space is a stretch that was achieved by Charles Howard Hinton in the $19^{th}$ century (2). What are the properties of spaces having even higher dimensions? What is it like to live in a space with one hundred dimensions? Or a million dimensions? Or a space like our brain that has a million billion dimensions (the number of synapses between neurons)?

The first Neural Information Processing Systems (NeurIPS) Conference and Workshop took place at the Denver Tech Center in 1987 (Fig. 2). The 600 attendees were from a wide range of disciplines, including physics, neuroscience, psychology, statistics, electrical engineering, computer science, computer vision, speech recognition and robotics. But they all had something in common: they all worked on intractably difficult problems that were not easily solved with traditional methods and they tended to be outliers in their home disciplines. In retrospect, 33 years later, these misfits were pushing the frontiers of their fields into high-dimensional spaces populated by big data sets, the world we are living in today. As the president of the Foundation that organizes the annual NeurIPS conferences, I oversaw the remarkable evolution of a community that created modern machine learning. This conference has grown steadily and in 2019 attracted over 14,000 participants. Many intractable problems eventually became tractable, and today machine learning serves as a foundation for contemporary artificial intelligence (AI).

The early goals of machine learning were more modest than those of artificial intelligence. Rather than aim directly at general intelligence, machine learning started by attacking practical problems in perception, language, motor control, prediction and inference using learning from data as the primary tool. In contrast, early attempts in AI were characterized by low-dimensional algorithms that were handcrafted. However, this approach only worked for well controlled environments. For example, in Blocks World, all objects were rectangular solids, identically painted and in an environment with fixed lighting. These algorithms did not scale up to vision in the real world, where objects have complex shapes, a wide range of reflectances and lighting conditions are uncontrolled. The real world is high-dimensional and there may not be any low-dimensional model that can be fit to it (3). Similar problems were encountered with early models of natural languages based on symbols and syntax, which ignored the complexities of semantics (4). Practical natural language applications became possible once the complexity of deep learning language models approached the complexity of the real world. Models of natural language with millions of parameters and trained with millions of labeled examples are now used routinely. Even larger deep learning language networks are in production today, providing services to millions of users online, less than a decade since they were introduced.

**Origins of Deep Learning.** I have written a book, *The Deep Learning Revolution: Artificial Intelligence Meets Human Intelligence* (5), which tells the story of how deep learning came about. Deep learning was inspired by the massively parallel architecture found in brains and its origins can be traced to Frank Rosenblatt's perceptron (6) in the 1950s that was based on a simplified model of a single neuron introduced by McCulloch and Pitts (7). The perceptron performed pattern recognition and learned to classify labeled examples (Fig. 3). Rosenblatt proved a theorem that if there was a set of parameters that could classify new inputs correctly, and there were enough examples, his learning algorithm was guaranteed to find it. The learning algorithm used labeled data to make small changes to parameters, which were the weights on the inputs to a binary threshold unit, implementing gradient descent. This simple paradigm is at the core of much larger and more sophisticated neural network architectures today, but the jump from perceptrons to deep learning was not a smooth one. There are lessons to be learned from how this happened.

The perceptron learning algorithm required computing with real numbers, which digital computers performed inefficiently in the 1950s. Rosenblatt received a grant for the equivalent today of one million dollars from the Office of Naval Research to build a large analog computer that could perform the weight updates in parallel using banks of motor-driven potentiometers representing variable weights (Fig. 3). The great expectations in the press (Fig. 3) were dashed by Marvin Minsky and Seymour Papert, who showed in their book *Perceptrons* (8) that a perceptron can only represent categories that are linearly separable in weight space. Although, at the end of their book, Minsky and Papert considered the prospect of generalizing single- to multiple-layer perceptrons, one layer feeding into the next, they doubted there would ever be a way to train these more powerful multilayer perceptrons. Unfortunately, many took this doubt to be definitive, and the field was abandoned until a new generation of neural network researchers took a fresh look at the problem in the 1980s.

The computational power available for research in the 1960s was puny compared to what we have today; this favored programming rather than learning, and early progress with writing programs to solve toy problems looked encouraging. By the 1970s, learning had fallen out of favor, but by the 1980s digital computers had increased in speed making it possible to simulate modestly-sized neural networks. During the ensuing neural network revival in the 1980s, Geoffrey Hinton and I introduced a learning algorithm for Boltzmann machines proving that contrary to general belief, it was possible to train multilayer networks (9). The Boltzmann machine learning algorithm is local and only depends on correlations between the inputs and outputs of single neurons, a form of Hebbian plasticity that is found in the cortex (10). Intriguingly, the correlations computed during training must be normalized by correlations that occur without inputs, which we called the sleep state, to prevent self-referential learning. It is also possible to learn the joint probability distributions of inputs without labels in an unsupervised learning mode. However, another learning algorithm introduced at around the same time based on the backpropagation of errors was much more efficient, though at the expense of locality (11). Both of these learning algorithm use stochastic gradient descent, an optimization technique that incrementally changes the parameter values to minimize a loss function. Typically this is done after averaging the gradients for a small batch of training examples.

**Lost in Parameter Space.** The network models in the 1980s rarely had more than one layer of hidden units between the inputs and outputs, but they were already highly over-parameterized by the standards of statistical learning. Empirical studies uncovered a number of paradoxes that could not be explained at the time. Even though the networks were tiny by today's standards, they had orders of magnitude more parameters than traditional statistical models. According to bounds from theorems in statistics, generalization should not be possible with the relatively small training sets that were available. But even simple methods for regularization, such as weight decay, led to models with surprisingly good generalization.

Even more surprising, stochastic gradient descent of non-convex loss functions were rarely trapped in local minima. There were long plateaus on the way down when the error hardly changed, followed by sharp drops. Something about these network models and the geometry of their high-dimensional parameter spaces allowed them to navigate efficiently to solutions and achieve good generalization, contrary to the failures predicted by conventional intuition.

Network models are high-dimensional dynamical systems that learn how to map input spaces into output spaces. These functions have special mathematical properties that we are just beginning to understand. Local minima during learning are rare because in the high-dimensional parameter space, most critical points are saddle points (12). Another reason why good solutions can be found so easily by stochastic gradient descent is that, unlike low-dimensional models where a unique solution is sought, different networks with good performance converge from random starting points in parameter space. Because of over-parameterization (13), the degeneracy of solutions changes the nature of the problem from finding a needle in a haystack to a haystack of needles.

Many questions are left unanswered. Why is it possible to generalize from so few examples and so many parameters? Why is stochastic gradient descent so effective at finding useful functions compared to other optimization methods? How large is the set of all good solutions to a problem? Are good solutions related to each other in some way? What are the relationships between architectural features and inductive bias that can improve generalization? The answers to these questions will help us design better network architectures and more efficient learning algorithms.

What no one knew back in the 1980s was how well neural network learning algorithms would scale with the number of units and weights in the network. Unlike many AI algorithms that scale combinatorially, as deep learning networks expanded in size, training scaled linearly with the number of parameters and performance continued to improve as more layers were added (14). Furthermore, the massively parallel architectures of deep learning networks can be efficiently implemented by multicore chips. The complexity of learning and inference with fully parallel hardware is $O(1)$. This is a rare conjunction of favorable computational properties.

When a new class of functions is introduced, it takes generations to fully explore them. For example, when Joseph Fourier introduced Fourier series in 1807, he could not prove convergence and their status as functions was questioned. This did not stop engineers from using Fourier series to solve the heat equation and apply them to other practical problems. The study of this class of functions eventually led to deep insights into functional analysis, a jewel in the crown of mathematics.

**The Nature of Deep Learning.** The third wave of exploration into neural network architectures, unfolding today, has greatly expanded beyond its academic origins, following the first two waves spurred by perceptrons in the 1950s and multilayer neural networks in the 1980s. The press has rebranded deep learning as AI. What deep learning has done for AI is to ground it in the real world. The real world is analog, noisy, uncertain and high dimensional, which never jived with the black and white world of symbols and rules in traditional AI. Deep learning provides an interface between these two worlds. For example, natural language processing has traditionally been cast as a problem in symbol processing. However, end-to-end learning of language translation in recurrent neural networks extracts both syntactic and semantic information from sentences. Natural language applications often start not with symbols, but with word embeddings in deep learning networks trained to predict the next word in a sentence (15), which are semantically deep and represent relationships between words as well as associations. Once regarded as "just statistics," deep recurrent networks are high-dimensional dynamical systems through which information flows much as electrical activity flows through brains.

One of the early tensions in artificial intelligence research in the 1960s was its relationship to human intelligence. The engineering goal of artificial intelligence was to reproduce the functional capabilities of human intelligence by writing programs based on intuition. I once asked Allen Newell, a computer scientist from Carnegie Mellon University and one of the pioneers of artificial intelligence who attend the seminal Dartmouth summer conference in 1956, why AI pioneers had ignored brains, the substrate of human intelligence. The performance of brains was the only existence proof that any of the hard problems in AI could be solved. He told me that he personally had been open to insights from brain research, but there simply hadn't been enough known about brains at the time to be of much help.

Over time, the attitude in AI had changed from "not enough is known" to "brains are not relevant." This view was commonly justified by an analogy with aviation: "If you want to build a flying machine, you would be wasting your time studying birds that flap their wings or the properties of their feathers." Quite to the contrary, the Wright Brothers were keen observers of gliding birds, which are highly efficient flyers (16). What they learned from birds was ideas for designing practical airfoils and basic principles of aerodynamics. Modern jets have even sprouted winglets at the tips of wings, which saves 5% on fuel and look suspiciously like wingtips on eagles (Fig. 4). Much more is now known about how brains process sensory information, accumulate evidence, make decisions and plan future actions. Deep learning was similarly inspired by nature. There is a burgeoning new field in computer science, called algorithmic biology, which seeks to describe the wide range of problem-solving strategies used by biological systems (17). The lesson here is we can learn from nature general principles and specific solutions to complex problems, honed by evolution and passed down the chain of life to humans.

There is a stark contrast between the complexity of real neurons and the simplicity of the model neurons in neural network models. Neurons are themselves complex dynamical systems with a wide range of internal time scales. Much of the complexity of real neurons is inherited from cell biology – the need for each cell to generate its own energy and maintain homeostasis under a wide range of challenging conditions. But other features of neurons are likely to be important for their computational function, some of which have not yet been exploited in model networks. These features include a diversity of cell types, optimized for specific functions; short-term synaptic plasticity, which can be either facilitating or depressing on a time scales of seconds; a cascade of biochemical reactions underlying plasticity inside synapses controlled by the history of inputs that extends from seconds to hours; sleep states during which a brain goes offline to restructure itself; and communication networks that control traffic between brain areas (18). Synergies between brains and AI may now be possible that could benefit both biology and engineering.

The neocortex appeared in mammals 200 million years ago. It is a folded sheet of neurons on the outer surface of the brain, called the gray matter, which in humans is about 30 cm in diameter and 5 mm thick when flattened. There are about 30 billion cortical neurons forming 6 layers that are highly interconnected with each other in a local stereotyped pattern. The cortex greatly expanded in size relative the central core of the brain during evolution, especially in humans where it constitutes 80% of the brain volume. This expansion suggests that the cortical architecture is scalable-- more is better--unlike most brain areas, which have not expanded relative to body size. Interestingly, there are many fewer long-range connections than local connections, which form the white matter of the cortex, but its volume scales as the 5/4 power of the gray matter volume and becomes larger than the volume of the gray matter in large brains (19). Scaling laws for brain structures can provide insights into important computational principles (20). Cortical architecture including cell types and their connectivity is similar throughout the cortex, with specialized regions for different cognitive systems. For example, the visual cortex has evolved specialized circuits for vision, which have been exploited in convolutional neural networks (CNN), the most successful deep learning architecture. Having evolved a general purpose learning architecture, the neocortex greatly enhances the performance of many special purpose subcortical structures.

Brains have eleven orders of magnitude of spatially structured computing components (Fig. 5). At the level of synapses, each cubic millimeter of the cerebral cortex, about the size of a rice grain, contains a billion synapses. The largest deep learning networks today are reaching a billion weights. The cortex has the equivalent power of hundreds of thousands of deep learning networks, each specialized for solving specific problems. How are all these expert networks organized? The levels of investigation above the network level organize the flow of information between different cortical areas, a system-level communications problem. There is much to be learned about how to organize thousands of specialized networks by studying how the global flow of information in the cortex is managed. Long-range connections within the cortex are sparse because they are expensive, both because of the energy demand needed to send information over a long distance and also because they occupy a large volume of space. A switching network routes information between sensory and motor areas that can be rapidly reconfigured to meet ongoing cognitive demands (18).

Another major challenge for building the next generation of AI systems will be memory management for highly heterogeneous systems of deep learning specialist networks. There is need to flexibly update these networks without degrading already learned memories; this is the problem of maintaining stable, lifelong learning (21). There are ways to minimize memory loss and interference between subsystems. One way is to be selective about where to store a new experiences. This occurs during sleep, when the cortex enters globally coherent patterns of electrical activity. Brief oscillatory events, known as sleep spindles, recur thousands of times during the night and are associated with the consolidation of memories. Spindles are triggered by the replay of recent episodes experienced during the day and are parsimoniously integrated into long-term cortical semantic memory (22, 23).

**The Future of Deep Learning.** Although the focus today on deep learning was inspired by the cerebral cortex, a much wider range of architectures is needed to control movements and vital functions. Subcortical parts of mammalian brains essential for survival can be found in all vertebrates, including the basal ganglia that is responsible for reinforcement learning and the cerebellum, which provides the brain with forward models of motor commands. Humans are hypersocial, with extensive cortical and subcortical neural circuits to support complex social interactions (24). These brain areas will provide inspiration to those who aim to build autonomous AI systems.

For example, the dopamine neurons in the brainstem compute reward prediction error, which is a key computation in the temporal difference learning algorithm in reinforcement learning and, in conjunction with deep learning, powered AlphaGo to beat Ke Jie, the world champion Go player in 2017 (25). Recordings from dopamine neurons in the midbrain, which project diffusely throughout the cortex and basal ganglia, modulate synaptic plasticity and provide motivation for obtaining long-term rewards (26). Subsequent confirmation of the role of dopamine neurons in humans has led to a new field, neuroeconomics, whose goal is better understand how humans make economic decisions (27). Several other neuromodulatory systems also control global brain states to guide behavior, representing negative rewards, surprise, confidence and temporal discounting (28).

Motor systems are another area of AI where biologically-inspired solutions may be helpful. Compare the fluid flow of animal movements to the rigid motions of most robots. The key difference is the exceptional flexibility exhibited in the control of high-dimensional musculature in all animals. Coordinated behavior in high-dimensional motor planning spaces is an active area of investigation in deep learning networks (29). There is also a need for a theory of distributed control to explain how the multiple layers of control in the spinal cord, brainstem and forebrain are coordinated. Both brains and control systems have to deal with time delays in feedback loops, which can become unstable. The forward model of the body in the cerebellum provides a way to predict the sensory outcome of a motor command, and the sensory prediction errors are used to optimize open loop control. For example, the vestibulo-ocular reflex (VOR) stabilizes image on the retina despite head movements by rapidly using head acceleration signals in an open loop; the gain of the VOR is adapted by slip signals from the retina, which the cerebellum uses to reduce the slip (30). Brains have additional constraints due to the limited bandwidth of sensory and motor nerves, but these can be overcome in layered control systems with components having a diversity of speed-accuracy tradeoffs (31). A similar diversity is also present in engineered systems, allowing fast and accurate control despite having imperfect components (32).

**Towards artificial general intelligence.** Is there a path from the current state-of-the-art in deep learning to artificial general intelligence? From the perspective of evolution, most animals can solve problems needed to survive in their niches, but general abstract reasoning emerged more recently in the human lineage. However, we are not very good at it and need long training to achieve the ability to reason logically. This is because we are using brain systems to simulate logical steps that have not been optimized for logic. Students in grade school work for years to master simple arithmetic, effectively emulating a digital computer with a one second clock. Nonetheless, reasoning in humans is proof of principle that it should be possible to evolve large-scale systems of deep learning networks for rational planning and decision making. However, a hybrid solution might also be possible, similar to neural Turing machines developed by DeepMind for learning how to copy, sort, and navigate (33). According to Orgel's Second Rule, nature is cleverer than we are, but improvements may still be possible.

Recent successes with supervised learning in deep networks have led to a proliferation of applications where large data sets are available. Language translation was greatly improved by training on large corpora of translated texts. However, there are many applications for which large sets of labeled data are not available. Humans commonly make subconscious predictions about outcomes in the physical world, and are surprised by the unexpected. Self-supervised learning, in which the goal of learning is to predict the future output from other data streams is a promising direction (34). Imitation learning is also a powerful way learn important behaviors and gain knowledge about the world (35). Humans have many ways to learn and require a long period of development to achieve adult levels of performance.

Brains intelligently and spontaneously generate ideas and solutions to problems. When a subject is asked to lie quietly at rest in a brain scanner, activity switches from sensorimotor areas to a default mode network of areas that support inner thoughts, including unconscious activity. Generative neural network models can learn without supervision, with the goal of learning joint probability distributions from raw sensory data, which is abundant. The Boltzmann machine is an example of generative model (9). After a Boltzmann machine has been trained to classify inputs, clamping an output unit on generates a sequence of examples from that category on the input layer (36). Generative adversarial networks (GANs) can also generate new samples from a probability distribution learned by self-supervised learning (37). Brains also generate vivid visual images during dream sleep that are often bizarre.

**Looking ahead**. We are at the beginning of a new era that could be called the age of information. Data are gushing from sensors, the sources for pipelines that turn data into information, information into knowledge, knowledge into understanding, and, if we are fortunate, knowledge into wisdom. We have taken our first steps toward dealing with complex high-dimensional problems in the real world; like a baby's, they are more stumble than stride, but what is important is that we are heading in the right direction. Deep learning networks are bridges between digital computers and the real world; this allows us to communicate with computers on our own terms. We already talk to smart speakers, which will become much smarter. Keyboards will become obsolete, taking their place in museums alongside typewriters. This makes the benefits of deep learning available to everyone.

In his essay on "The unreasonable effectiveness of mathematics in the natural sciences," Eugene Wigner marveled that the mathematical structure of a physical theory often reveals deep insights into that theory that lead to empirical predictions (38). Also remarkable is that there are so few parameters in the equations, called physical constants. The title of this article mirrors Wigner's. But unlike the laws of physics, there is an abundance of parameters in deep learning networks and they are variable. We are just beginning to explore representation and optimization in very high-dimensional spaces. Perhaps someday an analysis of the structure of deep learning networks will lead to theoretical predictions and reveal deep insights into the nature of intelligence. We can benefit from the blessings of dimensionality.

Having found one class of functions to describe the complexity of signals in the world, perhaps there are others. Perhaps there is a universe of massively-parallel algorithms in high-dimensional spaces that we have not yet explored, which go beyond intuitions from the three-dimensional world we inhabit and the one-dimensional sequences of instructions in digital computers. Like the gentleman square in flatland (Fig. 1) and the explorer in the Flammarion engraving (Fig. 6), we have glimpsed a new world stretching far beyond old horizons.

**Figure Captions**

**Figure 1**. **Cover of the 1884 edition of *Flatland: A Romance in Many Dimensions* by Edwin A. Abbott (1).** Inhabitants were two-dimensional shapes, with their rank in society determined by the number of sides.

**Figure 2**. **The Neural Information Processing Systems conference brought together researchers from many fields of science and engineering**. The first Conference was held at the Denver Tech Center in 1987 and has been held annually since then. The first few meetings were sponsored by the IEEE Information Theory Society.

**Figure 3**. **Early perceptrons were large-scale analog systems (4).** (A) An analog perceptron computer receiving a visual input. The racks contained potentiometers driven by motors whose resistance was controlled by the perceptron learning algorithm (B) Article in the *New York Times*, July 8, 1958, from a UPI wire report. The perceptron machine was expected to cost $100,000 on completion in 1959, or around $1 million in today's dollars; the IBM 704 computer that cost $2 million in 1958, or $20 million in today's dollars, could perform 12,000 multiplies per second, which was blazingly fast at the time. The much less expensive Samsung Galaxy S6 phone, which can perform 34 billion operations per second, is more than a million times faster. Photo courtesy of George Nagy.

**Figure 4. Nature has optimized birds for energy efficiency**. (A) The curved feathers at the wingtips of an eagle boosts energy efficiency during gliding. (B) Winglets on a commercial jets save fuel by reducing drag from vortices.

**Figure 5. Levels of investigation of brains**. Energy efficiency is achieved by signaling with small numbers of molecules at synapses. Interconnects between neurons in the brain are three dimensional. Connectivity is high locally, but relatively sparse between distant cortical areas. The organizing principle in the cortex is based on multiple maps of sensory and motor surfaces in a hierarchy. The cortex coordinates with many subcortical areas to form the central nervous system (CNS) that generates behavior (Adapted from *The Computational Brain*, Churchland, P. and Sejnowski, T., MIT Press, 1992).

**Figure 6.** Engraving from Camille Flammarion's 1888 book *L'atmosphère: météorologie populaire* ("*The Atmosphere: Popular Meteorology*,") Paris: Hachette. *p. 163*. The caption that accompanies the engraving in Flammarion's book reads: "A missionary of the Middle Ages tells that he had found the point where the sky and the Earth touch …"

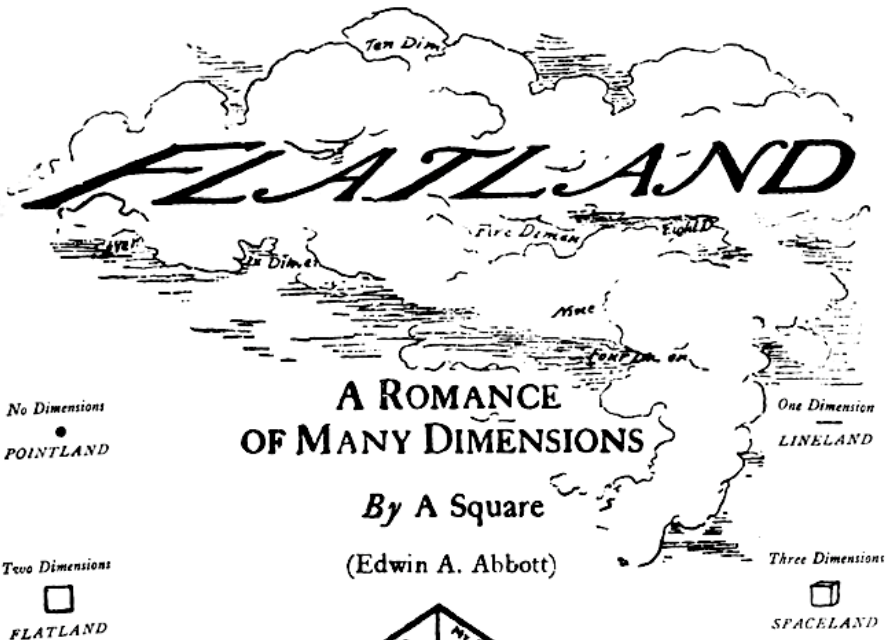

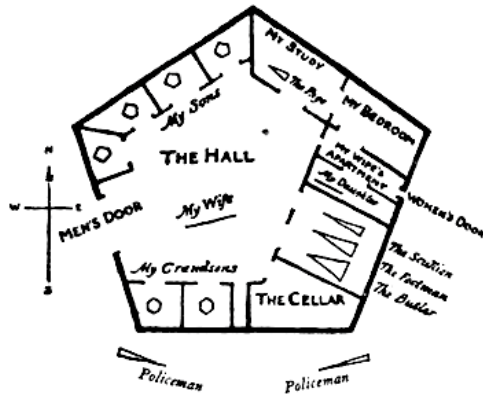

**Figure 1**. **Cover of the 1884 edition of *Flatland: A Romance in Many Dimensions* by Edwin A. Abbott (1).** Inhabitants were two-dimensional shapes, with their rank in society determined by the number of sides.

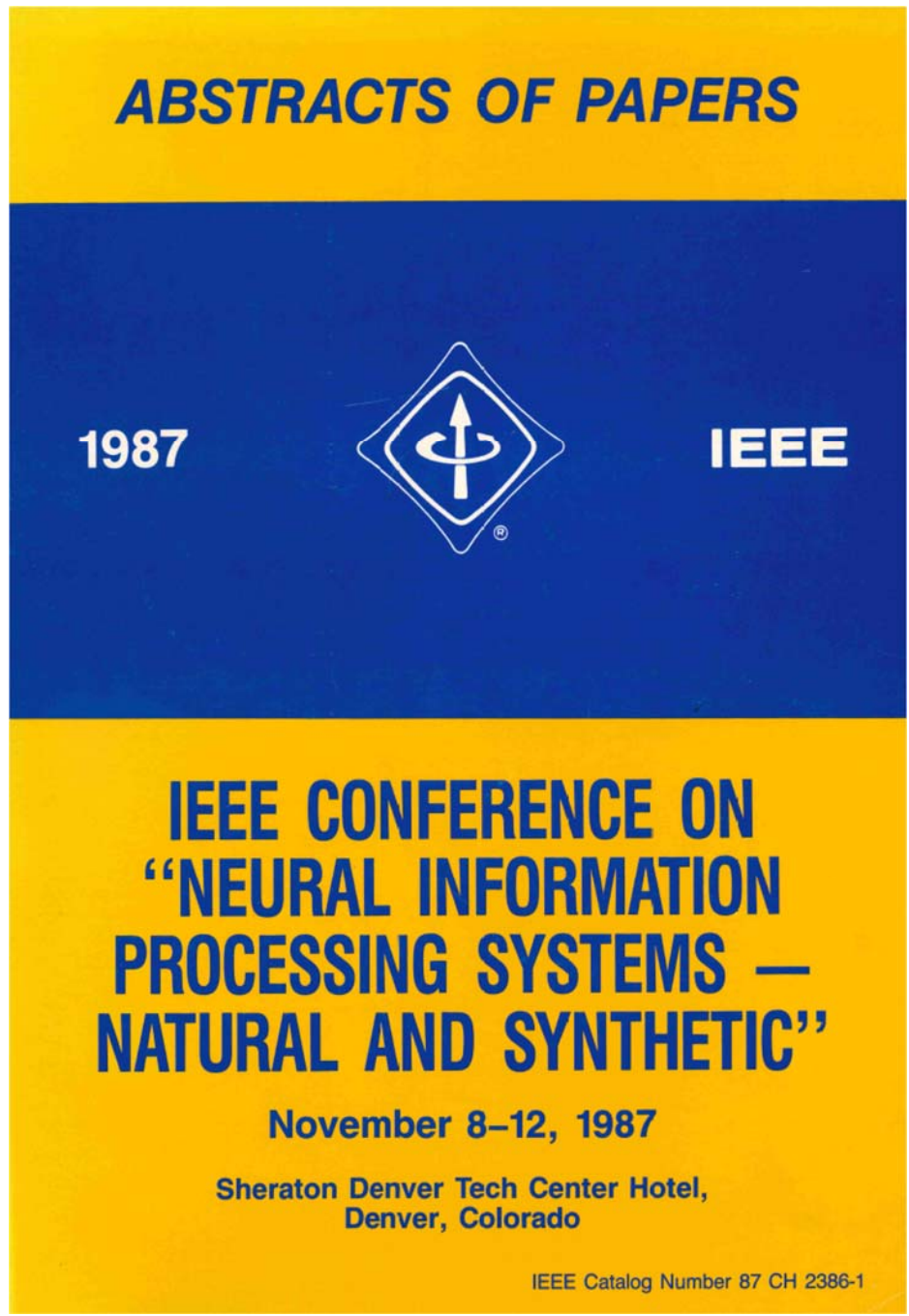

**Figure 2**. **The Neural Information Processing Systems conference brought together researchers from many fields of science and engineering**. The first Conference was held at the Denver Tech Center in 1987 and has been held annually since then. The first few meetings were sponsored by the IEEE Information Theory Society.

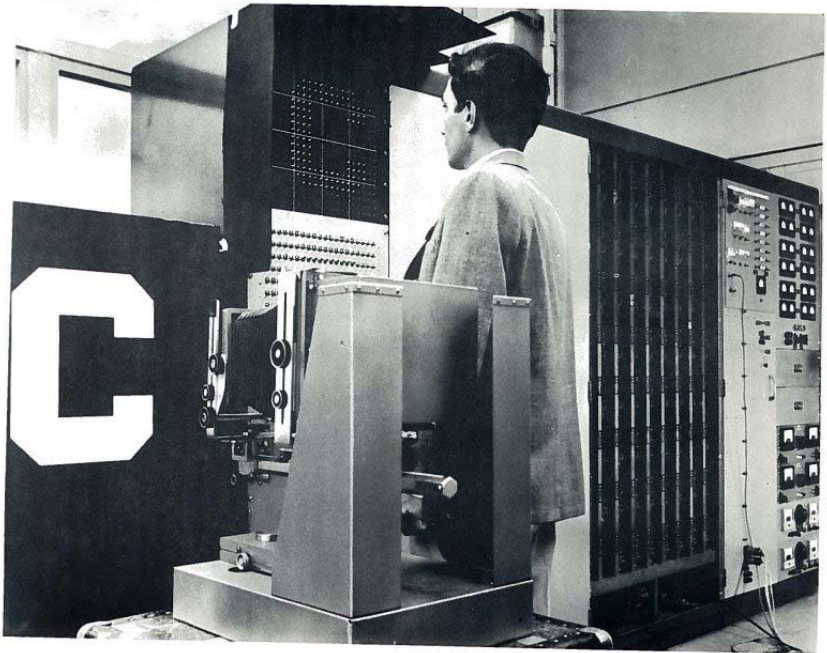

THE MARK I PERCEPTRON

**Figure 3**. **Early perceptrons were large-scale analog systems (4).** (Above) An analog perceptron computer receiving a visual input. The racks contained potentiometers driven by motors whose resistance was controlled by the perceptron learning algorithm. (from Ref 6) . (Right) Article in the *New York Times*, July 8, 1958, from a UPI wire report. The perceptron machine was expected to cost $100,000 on completion in 1959, or around $1 million in today's dollars; the IBM 704 computer that cost $2 million in 1958, or $20 million in today's dollars, could perform 12,000 multiplies per second, which was blazingly fast at the time. The much less expensive Samsung Galaxy S6 phone, which can perform 34 billion operations per second, is more than a million times faste.

## NEW NAVY DEVICE LEARNS BY DOING

Psychologist Shows Embryo of Computer Designed to Read and Grow Wiser

WASHINGTON, July 7 (UPI) —The Navy revealed the embryo of an electronic computer today that it expects will be able to walk, talk, see, write, reproduce itself and be conscious of its existence.

The embryo—the Weather Bureau's $2,000,000 "704" computer—learned to differentiate between right and left after fifty attempts in the Navy's demonstration for newsmen.

The service said it would use this principle to build the first of its Perceptron thinking machines that will be able to read and write. It is expected to be finished in about a year at a cost of $100,000.

Dr. Frank Rosenblatt, designer of the Perceptron, conducted the demonstration. He said the machine would be the first device to think as the human brain. As do human beings, Perceptron will make mistakes at first, but will grow wiser as it gains experience, he said.

Dr. Rosenblatt, a research psychologist at the Cornell Aeronautical Laboratory, Buffalo, said Perceptrons might be fired to the planets as mechanical space explorers.

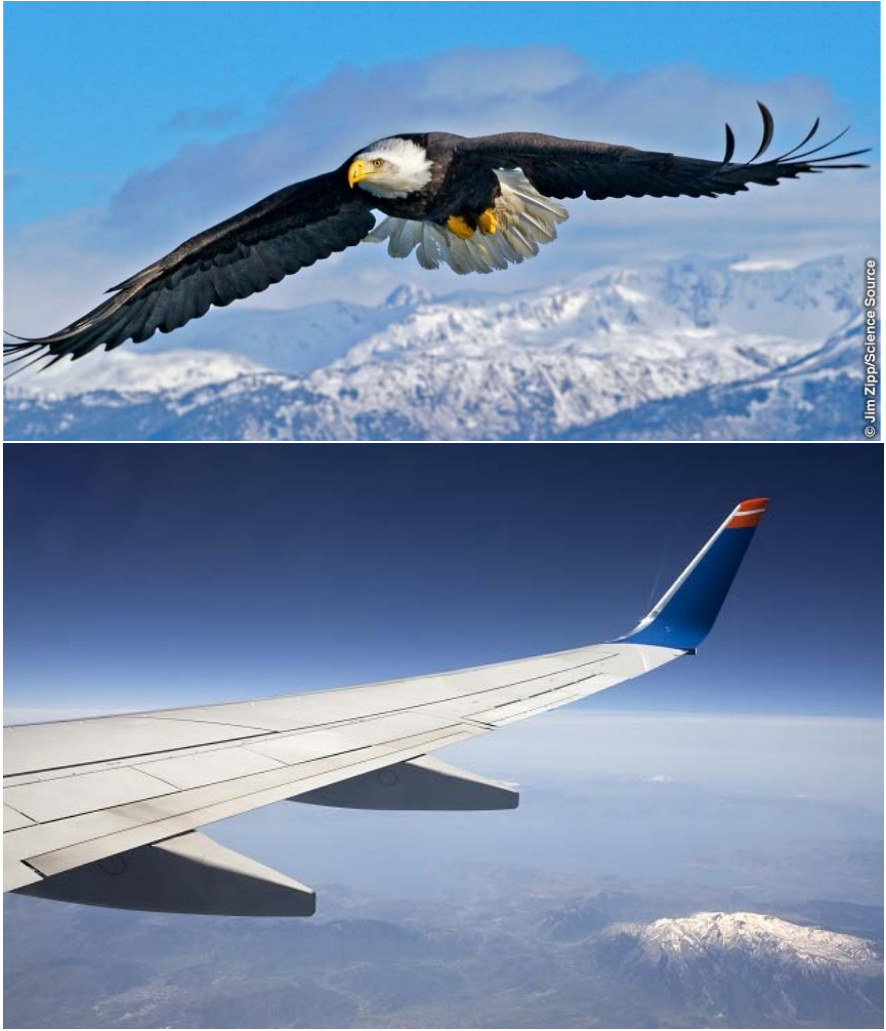

**Figure 4. Nature has optimized birds for energy efficiency**. (Top) The curved feathers at the wingtips of an eagle boosts energy efficiency during gliding. (Bottom) Winglets on a commercial jets save fuel by reducing drag from vortices.

# Levels of Investigation

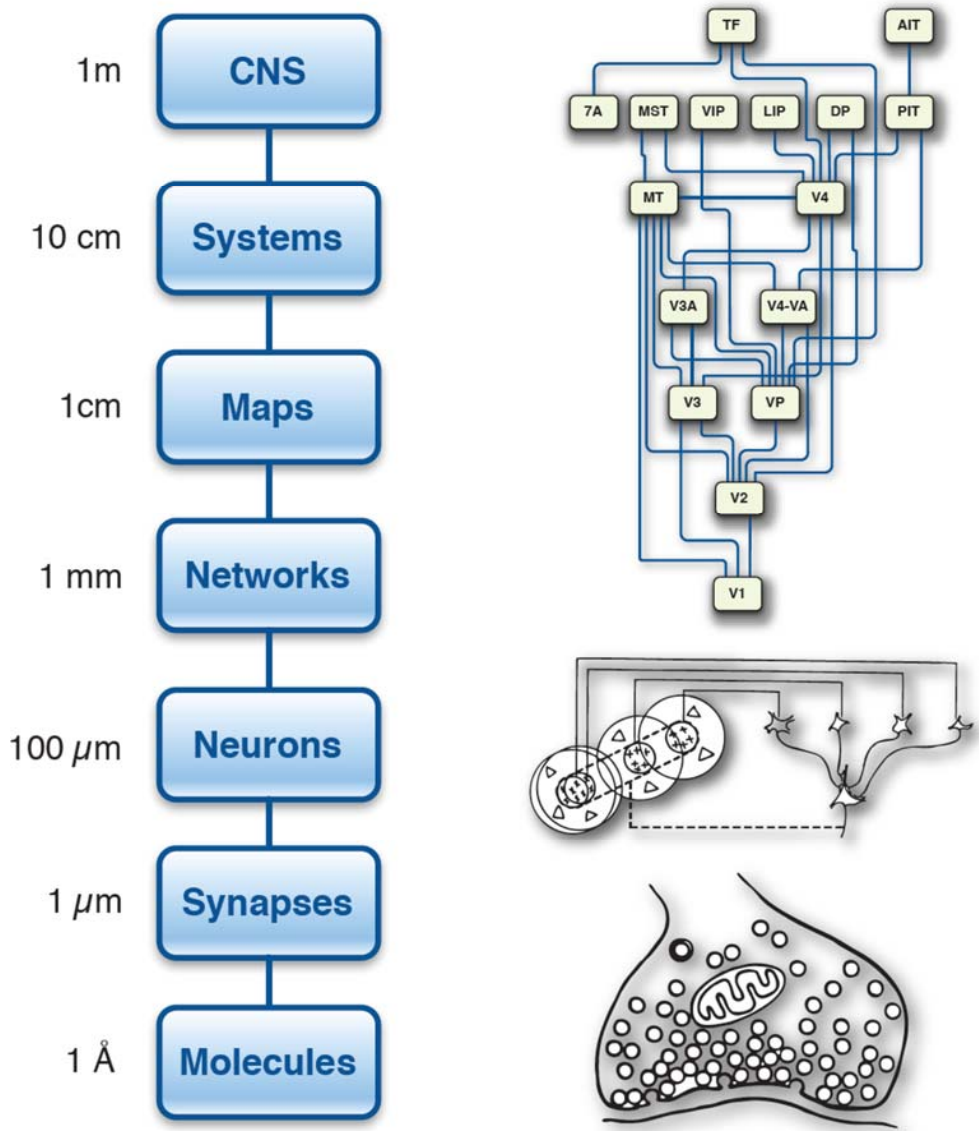

**Figure 5. Levels of investigation of brains**. Energy efficiency is achieved by signaling with small numbers of molecules at synapses. Interconnects between neurons in the brain are three dimensional. Connectivity is high locally, but relatively sparse between distant cortical areas. The organizing principle in the cortex is based on multiple maps of sensory and motor surfaces in a hierarchy. The cortex coordinates with many subcortical areas to form the central nervous system (CNS) that generates behavior (Adapted from *The Computational Brain*, Churchland, P. and Sejnowski, T., MIT Press, 1992).

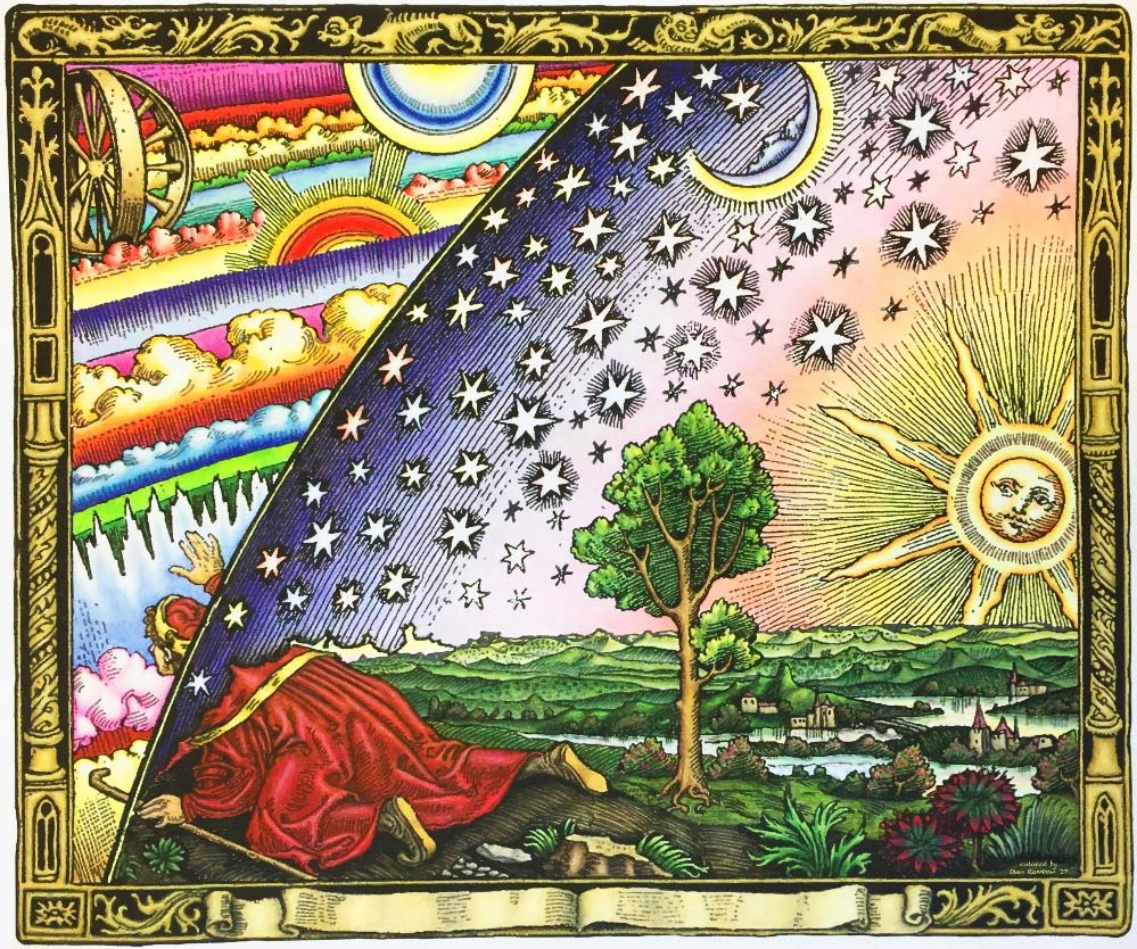

**Figure 6.** Engraving from Camille Flammarion's 1888 book *L'atmosphère: météorologie populaire* ("*The Atmosphere: Popular Meteorology*,") Paris: Hachette. *p. 163.* The caption that accompanies the engraving in Flammarion's book reads: "A missionary of the Middle Ages tells that he had found the point where the sky and the Earth touch …" Creative Commons License: https://commons.wikimedia.org/wiki/File:Flammarion_Colored.jpg